\documentstyle[epsfig]{nature_local}

\newcommand{\etal}{et al.}
\newcommand{\Casa}{Cas~A}
\newcommand{\zg}{z_{\mathrm{g}}}
\newcommand{\Mns}{M_{\mathrm{NS}}}
\newcommand{\Rns}{R_{\mathrm{NS}}}
\newcommand{\Tns}{T}
\newcommand{\Rem}{R}
\newcommand{\Teff}{T_{\mathrm{eff}}}
\newcommand{\Rinfty}{R^\infty}
\newcommand{\Tinfty}{T^\infty}
\newcommand{\NH}{N_{\mathrm{H}}}
\newcommand{\gtrsim}{\mathrel{\raise.3ex\hbox{$>$}\mkern-14mu
             \lower0.6ex\hbox{$\sim$}}}
\newcommand{\lesssim}{\mathrel{\raise.3ex\hbox{$<$}\mkern-14mu
             \lower0.6ex\hbox{$\sim$}}}

\title{A neutron star with a carbon atmosphere in the \\ Cassiopeia~A
supernova remnant}

\author{Wynn C. G. Ho
\affiliation{School of Mathematics, University of Southampton,
Southampton, SO17 1BJ, United Kingdom.}
\& Craig O. Heinke
\affiliation{Department of Physics, University of Alberta,
Room 238 CEB, 11322-89 Avenue, Edmonton, AB, T6G 2G7, Canada.}
}

\headertitle{Carbon Atmosphere NS in \Casa\ SNR}
\mainauthor{Ho \& Heinke}

\summary{\noindent
The surface of hot neutron stars is covered by a thin atmosphere.
If there is accretion after neutron star formation, the atmosphere could
be composed of light elements (H or He); if no accretion takes place
or if thermonuclear reactions occur after accretion, heavy elements
(for example, Fe) are expected.
Despite detailed searches, observations have been unable to confirm the
atmospheric composition of isolated neutron stars\cite{pavlovetal02}.
Here we report an analysis of archival observations of the compact
X-ray source in the centre of the Cassiopeia~A supernova remnant.
We show that a carbon atmosphere neutron star (with low magnetic field)
produces a good fit to the spectrum.  Our emission model, in contrast with
others\cite{pavlovetal00,chakrabartyetal01,pavlovluna09}, implies an
emission size consistent with theoretical predictions for the radius of
neutron stars.  This result suggests that there is nuclear burning in
the surface layers\cite{rosen68,changbildsten04} and also identifies the
compact source as a very young ($\approx 330$-year-old) neutron star.
}

\begin{document}
\maketitle

Cassiopeia~A is one of the youngest-known supernova remnants
in the Milky Way and is at a distance of $d=3.4^{+0.3}_{-0.1}$~kiloparsecs
(kpc) from the Earth\cite{reedetal95}.
The supernova that gave rise to the remnant may have been observed
by John Flamsteed in 1680 (ref.~\pcite{ashworth80}); the implied age
coincides with estimates made by studying the expansion of the
remnant\cite{fesenetal06}.
Although the supernova remnant is extremely well-studied, the central
compact source was only identified in first-light Chandra X-ray
observations\cite{tananbaum99}.
(We shall refer to the compact source as \Casa.)

\begin{figure}
\centerline{\psfig{file=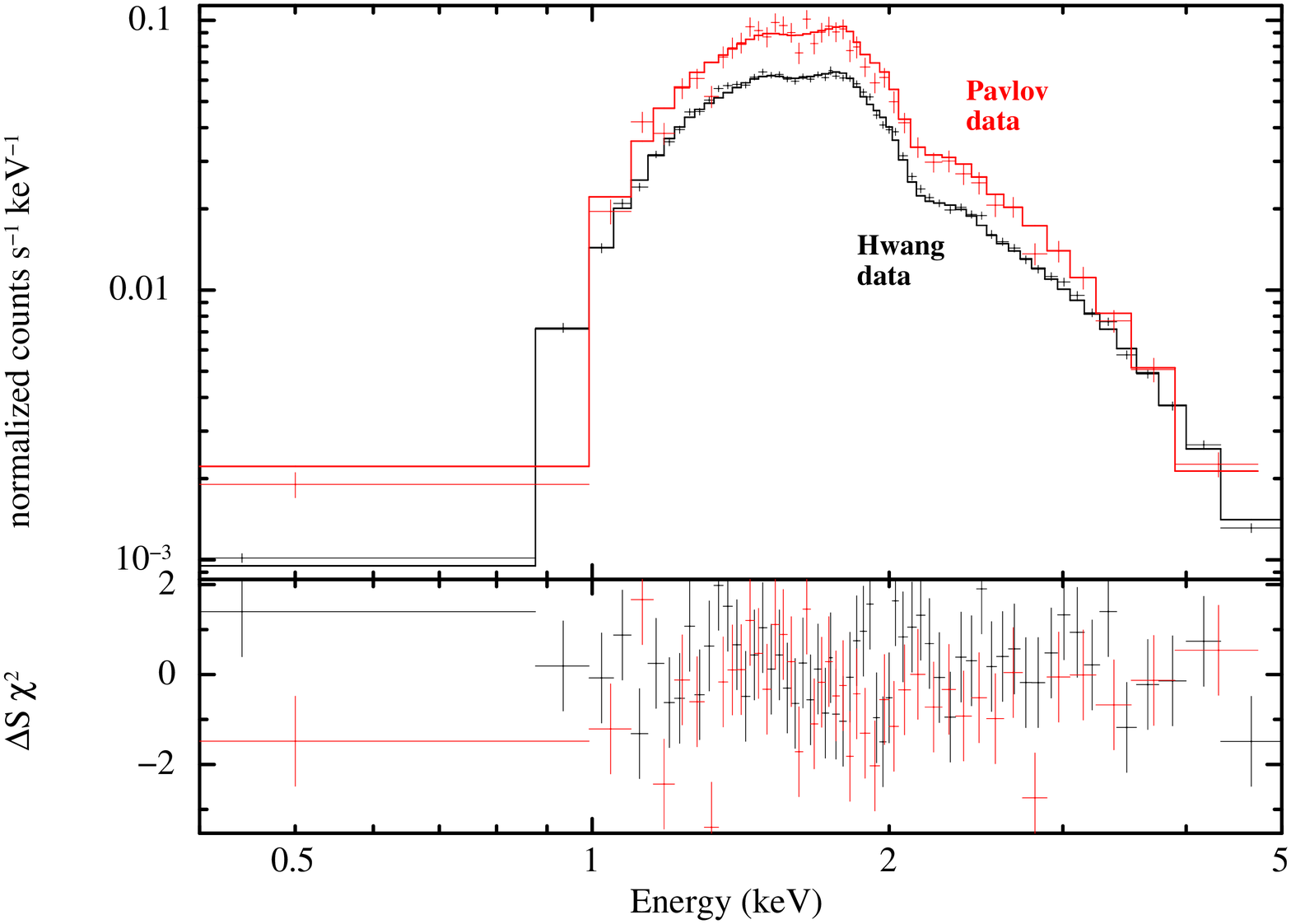,width=8.0cm}}
\caption{
{\bf Chandra X-ray spectra of \Casa.}
Spectra from the Hwang (black) and Pavlov (red) observations and fits with
our C spectral model.
Error bars are 1~s.d.
The lower panel shows the fit residuals $\Delta S_{\chi^2}$ in units of s.d.
The Hwang observations place \Casa\ off-axis (blurring Chandra's point-spread
function), and the high count-rate of \Casa\ distorts the spectrum through
detection of multiple photons during one 3.04-s frametime
(known as pile-up\cite{davis01}).
The Pavlov observation is performed in a special instrument configuration
to reduce the frametime to 0.34~s (thus reducing pile-up) and places
\Casa\ at the position of best focus\cite{pavlovluna09}.
We used the level~2 event files provided by the Chandra Data Archive and
performed data reduction and analysis with CIAO~4.1 and XSPEC~12.4.0.
Owing to the slightly distorted shape of the \Casa\ point-spread function
in the Hwang data, we extracted the source spectra using an elliptical
region of dimensions 1.23 by 1.72 arcsec, rotated to match the position
angle of the point-spread function ellipticity, and the background from a
surrounding annulus of 2.19 to 4.37 arcsec.
Alternative extraction methods produce similar results.
We combine the spectra and responses to make a single, deep spectrum of \Casa.
For the Pavlov data, we extract the \Casa\ spectrum and the nearby
background following the procedure of ref.~\pcite{pavlovluna09}.
The spectra are binned to achieve at least 1,000~counts per bin for the
Hwang data and 140~counts per bin for the Pavlov data.
X-ray and bolometric luminosities are
$L_X\mbox{(0.5-10 keV)}=4.3\times10^{33}$~ergs~s$^{-1}$ and
$L_{\rm bol}=7.0\times10^{33}$~ergs~s$^{-1}$, respectively.
\label{fig:obs}}
\end{figure}

We considered archival Chandra X-ray Observatory data from two studies of
\Casa, both using the ACIS-S charge-coupled device which provides spatial
and spectral information\cite{garmireetal03}.  A series of Chandra
observations, totalling 1~megasecond, was performed in 2004
to study the supernova remnant\cite{hwangetal04};
these are referred to here as the Hwang data.
A shorter (70~kiloseconds) observation in 2006 was designed to study the
compact source, here referred to as the Pavlov data\cite{pavlovluna09}.
Figure~\ref{fig:obs} shows the \Casa\ spectra, as well as our carbon model fit.

\begin{figure}
\centerline{\psfig{file=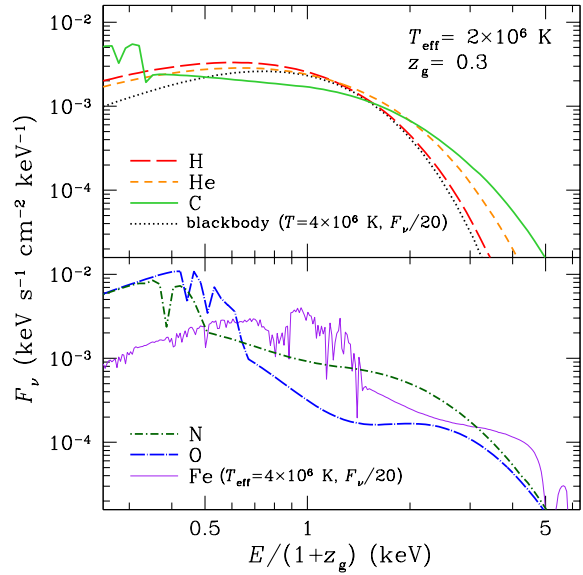,width=8.0cm}}
\caption{
{\bf Model atmosphere spectra.}
Energy flux $F_\nu$ for atmospheres with H, He, C, N, O, and Fe and a blackbody.
The energy has been redshifted by $1+\zg=(1-2G\Mns/c^2\Rns)^{-1/2}=1.3$,
where $\zg$ is the gravitational redshift,
and the flux has been scaled by $(10\mbox{ km}/3.4\mbox{ kpc})^2$.
Our models are constructed assuming a plane-parallel atmosphere
(because the atmosphere thickness of $\sim 1$~cm is much smaller than
the stellar size) that is in hydrostatic and radiative equilibrium
with constant gravitational acceleration $g=(1+\zg)\,G\Mns/\Rns^2$.
The efficient separation of light and heavy elements results in
atmospheres composed of a single element\cite{alcockillarionov80}.
The opacities are obtained
from tables computed by the Opacity Project\cite{opacity}.
(The energy range of the tables covers $E/kT\sim 0.07-20$.
When opacities are required beyond this range, we
use the $E^{-3}$-dependence of free-free and bound-free absorption;
this approximation has a minor effect, except for the Fe model, which
is shown for illustrative purposes.)
Light-element atmospheres generate spectra that are harder than
blackbodies (at the same temperature) because of the energy-dependence
of the opacity\cite{rajagopalromani96,zavlinetal96},
so atmospheric spectral fits result in temperatures that are lower and
sizes that are larger compared to those obtained using blackbody spectra.
Further details of the atmosphere model construction are given in
ref.~\pcite{holai01}.
When $B\ll 2.35\times 10^9Z^2$~G ($\sim 8\times10^{10}$~G for carbon),
magnetic-field effects on the radiation transport and atoms in the
atmospheric plasma are negligible\cite{rajagopalromani96,zavlinetal96}.
Previous works found poor fits to the data using magnetic ($B\ge 10^{12}$~G)
H atmosphere spectra\cite{pavlovetal00,pavlovluna09},
while magnetic mid-$Z$ element spectra are similar to low-magnetic field
Fe spectra in that they are blackbody-like in shape and contain many
lines\cite{moriho07}.
\label{fig:model}}
\end{figure}

We fitted the Hwang and Pavlov data simultaneously
with several models: a blackbody, a H atmosphere\cite{heinkeetal06}, and
atmospheres composed of pure He, C, N, or O;
these are illustrated in Fig.~\ref{fig:model}.
To identify promising models, the mass and radius of the atmosphere models
were fixed to the canonical neutron star values of $\Mns=1.4M_{\rm Sun}$ and
$\Rns=10$~km, where $M_{\rm Sun}$ is the solar mass.
The normalization factor, which can be interpreted as the fraction of the
surface that is emitting X-ray radiation, was left free.
The results are given in Table~\ref{tab:fit}.

The H, He, and C atmospheres provided good fits
to the data (somewhat better than the blackbody).
However, the derived sizes of the emission region $\Rem$ for H and He
(4~km and 5~km, respectively) are much smaller than the theoretical
size of a neutron star $\Rns$ ($\sim 10$~km; ref.~\pcite{lattimerprakash07}).
This would suggest the emission region is a hot spot on the neutron
star surface, which would probably result in X-ray pulsations as the
hot spot rotated with the star.  However, these pulsations have not
been detected\cite{pavlovluna09,murrayetal02}.
Fits to the data were performed using an additional (temperature)
component, for example, a second blackbody or atmosphere spectrum, which
produced inferred $\Rinfty$ of $\approx 0.2$ and 2~km for blackbody fits
or 0.4 and 11 km for H atmosphere fits\cite{pavlovluna09}.
The high-temperature component has been interpreted as being due to a small
hot polar cap, while the low-temperature component is due to emission
from the remaining cooler neutron star surface.
However, it is difficult to generate, through anisotropic heat conduction,
such small regions of high-temperature contrast on the neutron star
surface\cite{greensteinhartke83}.
Also, the two temperature regions would again probably produce (undetected)
X-ray pulsations.
On the other hand, by considering a C atmosphere, we derived an emission
size $\Rem\approx 12-15$~km that closely matches the theoretical
prediction for the radius of a neutron star
$\Rns\sim 10-14$~km\cite{lattimerprakash07}.
In this case, the X-ray observations are detecting emission from the
entire stellar surface, and therefore the emission would not necessarily
vary as the star rotates (although local temperature differences could
cause small brightness fluctuations on top of the bright background).
Thus we conclude that \Casa\ is consistent with a low-magnetic field
carbon atmosphere neutron star of mass $1.4 M_{\rm Sun}$, radius 12--15~km,
and surface temperature $\Tns=1.8\times 10^6$~K.

\begin{figure}
\centerline{\psfig{file=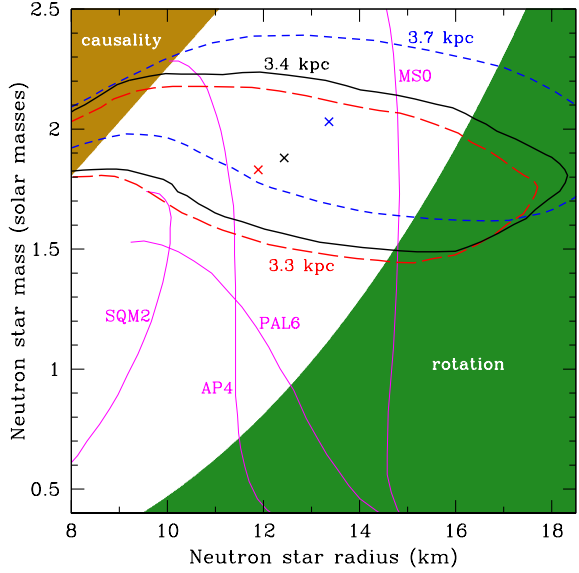,width=8.0cm}}
\caption{
{\bf Neutron star mass and radius.}
90\% confidence contours of $\chi^2$ around the best-fitting
\Casa\ mass and radius (crosses) for distances of 3.3~kpc
(red long-dashed line), 3.4~kpc (thick black solid line),
and 3.7~kpc (blue short-dashed line).
The upper-left and lower-right regions are excluded by constraints
from the requirement of causality and from the fastest-rotating neutron
star known\cite{lattimerprakash07}.
The thin magenta solid lines represent predictions for the masses and
radii of neutron stars using different theoretical models
(labelled SQM2, AP4, PAL6, and MS0) of the neutron star
interior\cite{lattimerprakash07}.
\label{fig:mr}}
\end{figure}

Interpreting the size of the emission region to be the true neutron
star radius ($\Rem=\Rns$),
we can constrain the neutron star mass $\Mns$ and radius $\Rns$ by
using a range of surface gravity models.
If we fix $\Mns=1.4M_{\rm Sun}$ and the distance to \Casa\ as
$d=3.4$~kpc (ref.~\pcite{reedetal95}), then we find
a surface effective temperature $\Teff=1.61^{+0.14}_{-0.05}\times 10^6$~K and
emission size $\Rem=15.6^{+1.3}_{-2.7}$~km.
Figure~\ref{fig:mr} shows 90\% confidence contours in mass and radius
when both are allowed to vary.
The contours for distances between 3.3 and 3.7~kpc constrain
$\Mns\approx 1.5-2.4M_{\rm Sun}$ and $\Rns\approx 8-17$~km.
The mass constraint, significantly above the canonical $1.4M_{\rm Sun}$,
suggests a moderately stiff nuclear equation of state\cite{lattimerprakash07}.

The emission of neutrinos determines the thermal evolution of a neutron star
at ages $\lesssim 10^5$~y.  The particle reactions that contribute to
this emission are determined by the state of the matter, which is strongly
dependent on the total mass of the star\cite{yakovlevpethick04,pageetal06}.
Owing to the small uncertainties in the temperature and age of \Casa,
as well as our measured mass, \Casa\ can be used to constrain matter
properties in the stellar interior.
More importantly, because the next youngest neutron stars for which surface
thermal emission has been measured have ages
exceeding a few thousand years, \Casa\ (with an age of only about 330~years)
serves as a valuable window into the early life of a passively-cooling
neutron star.
Previously, the utility of \Casa\ in studying the thermal evolution
was hindered by the fact that the measured temperature (taken to be
from a local hot spot because $\Rem\ll\Rns$) could only be used to set an upper
limit on the average surface temperature\cite{pavlovetal00}.
However, with our $\Teff$ and $L_{\rm bol}$ (which are surface averages
given that $\Rem=\Rns$), detailed comparisons with theoretical
models\cite{yakovlevpethick04,pageetal06,tsurutaetal09} can now be made.

The presence of a carbon atmosphere on the surface is probably a consequence
of the youth of \Casa.
The surface of newly-formed neutron stars is uncertain; it is
thought to be composed of an element between O and Fe, depending on which
layers of
the pre-supernova star fall back onto the proto-neutron star\cite{woosley+02};
this fallback material could also form light elements through
spallation\cite{changbildsten04}.
Over time, the surface accumulates an overlying layer of light
elements by accretion from the circumstellar medium
(although this can be countered by pulsar-wind
excavation\cite{changbildsten04}).
Fits to the thermal radiation from older neutron stars do indeed
suggest H or He atmospheres at $10^{4}-10^{5}$~years and heavy-element
(blackbody-like) atmospheres at $\gtrsim 10^5$~years\cite{pavlovetal02}.
However, the picture is incomplete for young neutron stars
($t\lesssim 10^4$~y).
At a depth just below the surface, the temperature (and density) is
conducive to nuclear burning ($T\gtrsim 10^8$~K).
Accreted material diffuses to this layer and is rapidly consumed,
with depletion of H in $\lesssim 1$~y and He in
$\lesssim 100$~y\cite{rosen68,changbildsten04}.
This process of diffusive nuclear burning is very temperature-sensitive;
as the neutron star cools, the temperature drops below the threshold
needed for the material to be burned, and a light-element atmosphere
can begin to build up after $10^4-10^6$~y (ref.~\pcite{changbildsten04}).

\Casa\ is the youngest of a class of neutron stars that are located
near the centre of supernova remnants, possess steady long-term fluxes
and soft thermal spectra, and have no detectable pulsar wind nebulae
or radio pulsations\cite{pavlovetal02}.
X-ray pulsations have been detected in three members of this class,
with periods of tenths of seconds.
Timing measurements suggest that they have (dipolar) magnetic fields
$B\ll 10^{12}$~G (ref.~\pcite{gotthelfhalpern09}).
If the field of \Casa\ is $\sim (1-5)\times 10^{11}$~G, then a spectral
feature due to the electron cyclotron resonance may appear in the
Chandra energy range, but this has yet to be detected.
The lack of a visible pulsar wind nebula and no indication of
magnetospheric activity (such as radio or gamma-ray emission or a
high-energy hard power-law component), like those seen in classical
pulsars (with $B\gtrsim 10^{12}$~G), also suggest the magnetic field is low.
The weak surface magnetic field inferred for this class would
have important implications for the neutron star population.
These objects could be representative of the early life of neutron stars
before becoming classical pulsars.
In this case, either neutron star magnetic fields develop by a dynamo
mechanism, or else a strong field (produced during the collapse of the
progenitor star) is buried and has not yet emerged\cite{muslimovpage95}.
These objects could also form a distinct group of low magnetic field
neutron stars,
which never manifest pulsar-like behaviour; this would suggest a large
population of old, cool, unobserved sources.
Finally, we note that the detection of the hydrogen-like C edge at
about 0.45~keV (unredshifted),
in similar sources with less intervening gas and dust than \Casa\ has,
would not only provide further evidence for carbon atmospheres but also
give a measurement of the gravitational redshift.

\smallskip
\noindent {\small {\bf Acknowledgements.}
W.C.G.H. thanks N. Badnell, P. Chang, and D. Lai for discussions.
W.C.G.H. appreciates the use of the computer facilities at the Kavli
Institute for Particle Astrophysics and Cosmology.
W.C.G.H. acknowledges support from the Science and Technology Facilities
Council (STFC) in the United Kingdom.
C.O.H. acknowledges support from the Natural Sciences and Engineering
Research Council (NSERC) of Canada.
}

\smallskip
\noindent {\small {\bf Author Contributions}
W.C.G.H. calculated the new models and wrote the manuscript.
C.O.H. reduced the data, fitted the models to the data, and contributed
to the manuscript.
}

\medskip
\noindent {\small {\bf Author Information}
 Reprints and permissions information is available at
 www.nature.com/reprints.
 The authors declare no competing financial interests.
 Correspondence and requests for materials should be
 addressed to W.C.G.H. (wynnho@slac.stanford.edu)
 or C.O.H. (cheinke@phys.ualberta.ca).}

\clearpage
\begin{table}
\small{
\begin{tabular}{lccccc}
\hline
Atmosphere model & $\NH$ & $kT$ & Normalization & $\chi^2$/d.o.f.
 & Null hypothesis \\
 & ($10^{22}$ cm$^{-2}$) & (eV) & & & probability (\%) \\
\hline
Hydrogen  & 1.65$^{+0.04}_{-0.05}$ & 241$^{+7}_{-6}$ & 0.18$^{+0.03}_{-0.03}$ &
106.3/99 & 29 \\
Helium    & 1.62$^{+0.04}_{-0.05}$ & 228$^{+9}_{-8}$ & 0.22$^{+0.05}_{-0.04}$ &
112.4/99 & 17 \\
Carbon    & 1.73$^{+0.04}_{-0.04}$ & 155$^{+7}_{-6}$ & 1.84$^{+0.56}_{-0.42}$ &
105.3/99 & 31 \\
Nitrogen  & 1.37 & 172 & 1.18 & 388/99 & 0 \\
Oxygen    & 1.03 & 234 & 0.20  & 2439/99  & 0 \\
\hline \\ \hline
Blackbody model & 1.46$^{+0.04}_{-0.04}$ & $k\Tinfty=387^{+7}_{-6}$ eV
 & $\Rinfty=1.0^{+0.1}_{-0.1}$ km & 134.2/98 & 11 \\
\hline
\end{tabular}
\caption{
{\bf X-ray spectral fitting of \Casa.}
Joint fit results to the Hwang and Pavlov data with
neutron star atmosphere and blackbody models that are modified by
photoelectric absorption $\NH$ (using the T\"{u}bingen-Boulder absorption
routine TBABS with wilms model abundances in the X-ray spectral fitting
package XSPEC; see http://heasarc.gsfc.nasa.gov/docs/xanadu/xspec/index.html),
dust scattering\cite{predehletal03}, and corrections for pile-up\cite{davis01}
(the grade migration parameter of the pile-up algorithm is allowed to
float freely and is 0.36 for the best-fit C model).
Chandra data with such high statistics reveals systematic uncertainties
in the Chandra calibration\cite{heinkeetal06},
so we added a systematic uncertainty of 3\% in quadrature.
All parameter errors are given at 90\% confidence.
Errors are not reported when the reduced $\chi^2 > 2$.
The normalization refers to the fraction of the neutron star surface
emitting radiation (for a 10~km stellar radius and 3.4~kpc distance).
The null hypothesis probability is the probability
that one realization of the model fit to the data would have a reduced
$\chi^2$ greater than that obtained; less than 5\% indicates a poor fit.
$\Tinfty=T/(1+\zg)$ and $\Rinfty=\Rem(1+\zg)$ are the temperature and
radius measured by an observer at infinity, and d.o.f. is the number of
degrees of freedom.}
\label{tab:fit}
}
\end{table}

\end{document}